\documentclass[a4paper,twoside]{article}

\usepackage[english]{babel}
\usepackage{epsfig}
\usepackage{subcaption}
\usepackage{calc}
\usepackage{amssymb}
\usepackage{amstext}
\usepackage{amsmath}
\usepackage{amsthm}
\usepackage{multicol}
\usepackage{pslatex}
\usepackage{apalike}
\usepackage{algorithm2e}
\usepackage[para]{footmisc}
\usepackage{SCITEPRESS}
\usepackage{cite}
\usepackage{amsmath,amssymb,amsfonts}
\usepackage{algorithmic}
\usepackage{textcomp}
\usepackage{xcolor}
\usepackage{enumitem}
\usepackage{graphicx}
\usepackage{hyperref}
\usepackage{caption}
\usepackage{booktabs}
\usepackage{multirow}
\usepackage{stfloats}
\usepackage[flushleft]{threeparttable}
\usepackage{tablefootnote}
\usepackage{amssymb}% http://ctan.org/pkg/amssymb
\usepackage{pifont}% http://ctan.org/pkg/pifont
\usepackage{pgfplots}
\pgfplotsset{compat=1.9}
\usepgfplotslibrary{external}
\pgfplotsset{every axis legend/.append style={font=\footnotesize}}

\renewcommand{\footnotesize}{\scriptsize}

\begin{document}

\title{A Heuristic Approach to Localize CSS Properties for \newline Responsive Layout Failures}

\author{\authorname{Tasmia Zerin, B M Mainul Hossain and Kazi Sakib}
\affiliation{Institute of Information Technology, University of Dhaka, Dhaka, Bangladesh}
\email{\{bsse1128, mainul, sakib\}@iit.du.ac.bd}
}


\abstract{% Responsive Layout Failures (RLFs) typically arise from CSS properties that hinder proper layout behavior in different screen sizes. To find an accurate and effective solution for repairing RLFs, localization of those properties is necessary. In the literature, most of the works are focused on detection and repair rather than localizing the root cause of the problem. This makes the proposed solutions less effective as they do not generate accurate repairs as a developer would do. To identify the similar problematic properties that the developers manually localize, a heuristic approach is proposed. The approach first detects the RLFs existing in a webpage and their affected elements. Then it localizes the nearby HTML elements using RLF direction and relative alignment of the elements present in the RLF region. The involved CSS properties of those elements are then identified using a ranked search set of CSS properties, created by analyzing Quora and Stack Overflow queries. Finally, the element and property pairs are ranked based on their impact on RLFs. We implemented this approach into a tool called {\normalfont \textsc{LocaliCSS}} and evaluated it on a set of webpages using Top N Rank, MRR and P@K metrics. It accurately localizes 45.2\% to 92.86\% of RLFs for Top-1 to Top-7 respectively. The approach also achieves an MRR of 76\% and P@3 of 77.13\%. We asked a group of experienced front-end engineers to manually localize the detected RLFs. Their preferred properties were found in our suggested list for Top-1 to Top-7 by  42.86\% to 90.48\% respectively.

Responsive Layout Failures (RLFs) typically arise from CSS properties that hinder proper layout behavior in different screen sizes. To find an accurate and effective solution for repairing RLFs, localization of those problematic properties is necessary.  However, existing approaches only detect RLFs and apply broad CSS patches for them. The patches alter the entire layout without localizing the root cause of failure. To address this gap, we propose a heuristic approach to identify the specific CSS properties that developers would typically localize manually. The approach first detects the RLFs existing in a webpage and their affected elements. Next, it localizes the nearby HTML elements using RLF direction and relative alignment of the elements present in the RLF region. The involved CSS properties of those elements are then identified using a ranked search set of CSS properties, created by analyzing Quora and Stack Overflow queries. Finally, elements and their corresponding property pairs are ranked based on their impact on RLFs. We have implemented this approach into a tool called {\normalfont \textsc{LocaliCSS}} and evaluated it on a set of webpages using Top N Rank, MRR and P@K metrics. The tool achieved localization accuracy ranging from 45.2\% (Top-1) to 92.86\% (Top-7), with an MRR of 76\% and a P@3 of 77.13\%. Additionally, experienced front-end engineers manually localized the RLFs as part of our evaluation. Their preferred CSS properties matched the suggestions from our approach in 42.86\% of cases for Top-1 rankings and up to 90.48\% for Top-7 rankings.}

\onecolumn \maketitle \normalsize \setcounter{footnote}{0} \vfill

\section{\uppercase{Introduction}}
\label{sec:introduction}
In current world, developing websites responsive to all screen sizes has gained popularity. This approach is known as \textit{Responsive Web Design} (RWD), which allows webpages to render well across all resolutions and screen sizes \cite{automated_detection}. However, ensuring responsiveness with the correct layout is challenging, often results in visual failures, known as \textit{Responsive Layout Failures} (RLFs) \cite{automated_detection}. RLFs usually occur due to insufficient space for webpage elements to render properly at a specific screen size or across a range of sizes. These failures include, HTML elements being cropped off the edge of the screen or colliding over one another and overwriting each other’s content. RLFs can occur into designs and may remain unnoticed until webpages go live, often because they occur at a very small range of viewport sizes out of a very wide range. Although many modern frameworks can now automatically make websites responsive, many existing websites and legacy platforms still rely on raw CSS styles, making RLFs more common.\\
The root cause of an RLF usually lies in the affected elements, but these elements may not always be the actual source of the failure. For instance, consider a row of elements \textit{e}, where the leftmost element is \textit{e1} and the rightmost element is \textit{e2}. The developer has added a large \texttt{\small margin-right} value to \textit{e1}, as a result \textit{e2} overflows its container. Here, while \textit{e2} is the affected element, actual root cause lies in the \texttt{\small margin-right} CSS property of \textit{e1}. A developer has to search or localize these properties manually, change and verify each of them whether they are the root cause.\\
Automating CSS localization can reduce this manual trial and error while searching. However, this localization should be effective enough to suggest properties similar to manual localization. At the same time, this approach needs to be efficient enough to accurately localize these elements and CSS properties responsible for RLFs.\\
In literature, approaches to detect and repair RLFs exist, e.g., ReDeCheck \cite{redecheck} detects RLFs of a webpage by comparing the layout in different viewports. Layout DR \cite{automated_repair} generates CSS patches as hotfixes for the detected RLFs. However, these patches modify CSS property values of the entire layout. Whereas, repairing only the RLF segment requires localizing the involved properties. While other web application failures are repaired through localization \cite{mob-friendly-problems, aesthetics}, no existing approach has yet worked on localizing the RLFs before repairing.\\
\looseness=-1
This paper presents a heuristic approach to localize HTML elements and CSS properties responsible for an RLF. We implemented this into a tool called {\normalfont \textsc{LocaliCSS}} (\underline{\textit{Locali}}zation of \underline{\textit{CSS}}). RLFs may occur at a small range of screen widths, known as \textit{viewports}. Hence, we start by detecting RLFs at different viewports across a webpage. Next, we find the involved elements by filtering surrounding elements of an RLF based on its type, direction and alignment. To localize the problematic properties of these elements, we created a ranked CSS property set for each RLF by analyzing Quora and Stack Overflow queries with expert's opinion. Lastly, each element and property pair is prioritized according to its impact on the failure. The impact is measured based on the property value and its rank in the CSS property set, with higher values and lower ranks indicating greater impact.\\
\looseness=-1
We applied {\normalfont \textsc{LocaliCSS}} to an existing set of 20 responsive webpages \cite{automated_repair, automated_detection}. Our empirical study exhibits the accuracy of this approach in identifying the correct element and property pairs using three evaluation metrics - Top N Rank, Mean Reciprocal Rank (MRR) and Precision at K (P@K). {\normalfont \textsc{LocaliCSS}} accurately localized and ranked the problematic pairs for 39 out of 42 detected RLFs. It achieved Top-1, Top-3, Top-5 and Top-7 accuracy in 45.2\%, 76.2\%, 90.5\% and 92.86\%  of cases respectively. For MRR, 68\% accuracy has been achieved by {\normalfont \textsc{LocaliCSS}}. After having experts opinions, some RLFs were marked as No Problems (NP). Excluding such cases gives us an MRR of 76\%. To measure relevance of the suggested properties, P@3 is calculated, which is 66.67\%. Excluding NP ones leads to a higher P@3 value of 77.12\%.\\
The detected RLFs were also provided to five experienced front-end engineers for manual localization. Our approach matches with their preferences for Top-1 by 42.86\%, which means their chosen property is ranked 1 in the suggested list by our approach. Similarly, for Top-3, 73.81\% of the time our suggested list contains the developer's choice in the top 3. Top-5 and Top-7 both give an accuracy of 90.48\%. These findings show that our proposed approach can effectively identify problematic CSS properties and can be used by a developer.

\section{\uppercase{Background}}
\label{sec:background}
\begin{figure*}
    \centering
    \includegraphics[width=\textwidth]{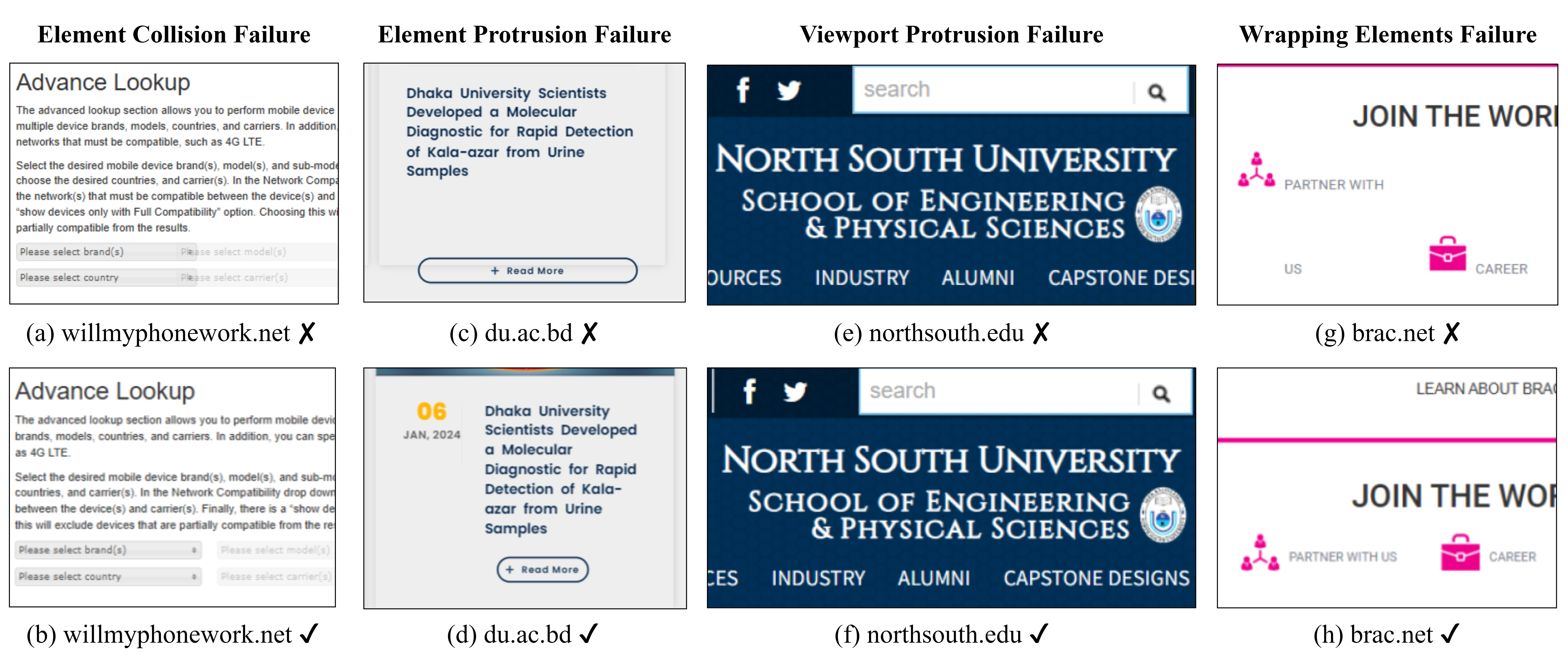}
    \caption{Real-life examples demonstrating RLFs and correct layouts}
    \label{fig:RLFs_comparison}
\end{figure*}

Responsive Web Design (RWD) is a design strategy to develop webpages so that the layout of the HTML elements automatically adjusts to the available space, allowing it to fit small mobiles to large desktop screens. Developers currently use this to develop responsive webpages. The available space or the width is known as a viewport, and this viewport width is considered from 320 pixels wide for a mobile device up to a wide width of 1400 pixels for desktop screens. RWD can be implemented using HTML and cascading style sheet (CSS) code or frameworks such as Bootstrap \footnote{\href{https://getbootstrap.com/}{Bootstrap}}, TailwindCSS \footnote{\href{https://tailwindcss.com}{Tailwind CSS}}, Bulma \footnote{\href{https://bulma.io.}{Bulma CSS}} and WindiCSS \footnote{\href{https://windicss.org/}{Windi CSS}}. These help developers create a responsive page to fit every viewport in a broad range of viewport sizes. \\
Despite these, Responsive Layout Failures (RLFs) are frequently occurring in the responsive webpages, as reported by Walsh et al. \cite{automated_detection}. RLFs are visual discrepancies in the rendering of a webpage which causes the webpage to deviate from its intended layout.  There are five common RLF types that Walsh et al. \cite{automated_detection} reported to be prevalent in real life. To demonstrate these failures, screenshots of responsively designed real-life web pages are shown in Figure 1. Left part of this figure (Figure \ref{fig:RLFs_comparison}(a), (c), (e), (g)) highlights each responsive layout failure, and the right part (Figure \ref{fig:RLFs_comparison}(b), (d), (f), (h)) highlights a correct layout of each of them respectively. These RLF types are explained below:\\
\textbf{Element Collision (EC).} Elements collide into one another due to insufficient accommodation space when viewport width reduces. Figure 1(a) shows an example of this type, where two buttons are colliding with one another. In a wider layout, the buttons are positioned correctly as shown in Figure 1(b).\\
\textbf{Element Protrusion (EP).} When the child element is contained within its container, but as the viewport width decreases, it lacks sufficient space to fit within its parent. As a result, the child element protrudes out of its container. In Figure 1(c) two buttons are getting out of their container, as a result, RLF occurs. When the viewport size is increased, the layout changes while buttons are inside their parent container.\\
\textbf{Viewport Protrusion (VP).} As the viewport size decreases, elements may not only overflow their containers but also protrude out of the viewable area of the webpage (i.e., the \texttt{<BODY>} tag), causing them to appear outside the horizontally visible portion of the page. In Figure 1(f), there are multiple options in a row. When the viewport size decreases, some options may overflow the viewport due to not having enough space. As a result, in Figure \ref{fig:RLFs_comparison}(f) some options are getting cropped on the right side of the webpage. \\
\textbf{Wrapping Elements.} When the container is not wide enough but has a flexible height, horizontally aligned elements contained within it no longer fit side by side, causing “wrap” to a new line on the page. The text of the first option of Figure 1(g) is wrapped to a new line, but in a wider viewport, the text stays in one line. Due to insufficient width space, this failure occurs.\\
\textbf{Small-Range Failure.} Responsively designed websites typically rely on numerous CSS rules that are triggered based on different media queries. In some cases, multiple media queries may be enabled simultaneously for a given viewport width. For example, if one rule applies when the viewport exceeds 768 pixels and another applies when it is below 1024 pixels, both rules will be active within the range of 769–1023 pixels. As it is only concerned to media-query rules rather than CSS properties, we have not considered this RLF type in this research.\\
RLFs are usually caused by one or multiple CSS properties of an HTML element, here the element can be one of the affected elements or any other element from the neighborhood. Localization can search all the probable problematic elements and their properties to find the ones that most likely need to be adjusted to resolve each RLF. From Figure \ref{fig:RLFs_comparison}(c), manual inspection may reveal that the protruding element is overflowing its container due to an extra space between the elements above it. Similarly for Figure \ref{fig:RLFs_comparison}(g), elements in the row have large space between them; as a result, the first element is wrapped to a new line. Such RLFs occur in a small range of viewports due to inappropriate values of CSS properties. However, these problematic CSS properties may not necessarily always be of the affected elements, they can be located in other elements. Hence, localization of these properties can play a significant role in finding the root cause of an RLF.

\vspace{-20pt}
\vspace{5pt}

\section{\uppercase{Approach}}
\begin{figure}[ht]
    \centering
    \includegraphics[width=0.95\columnwidth]{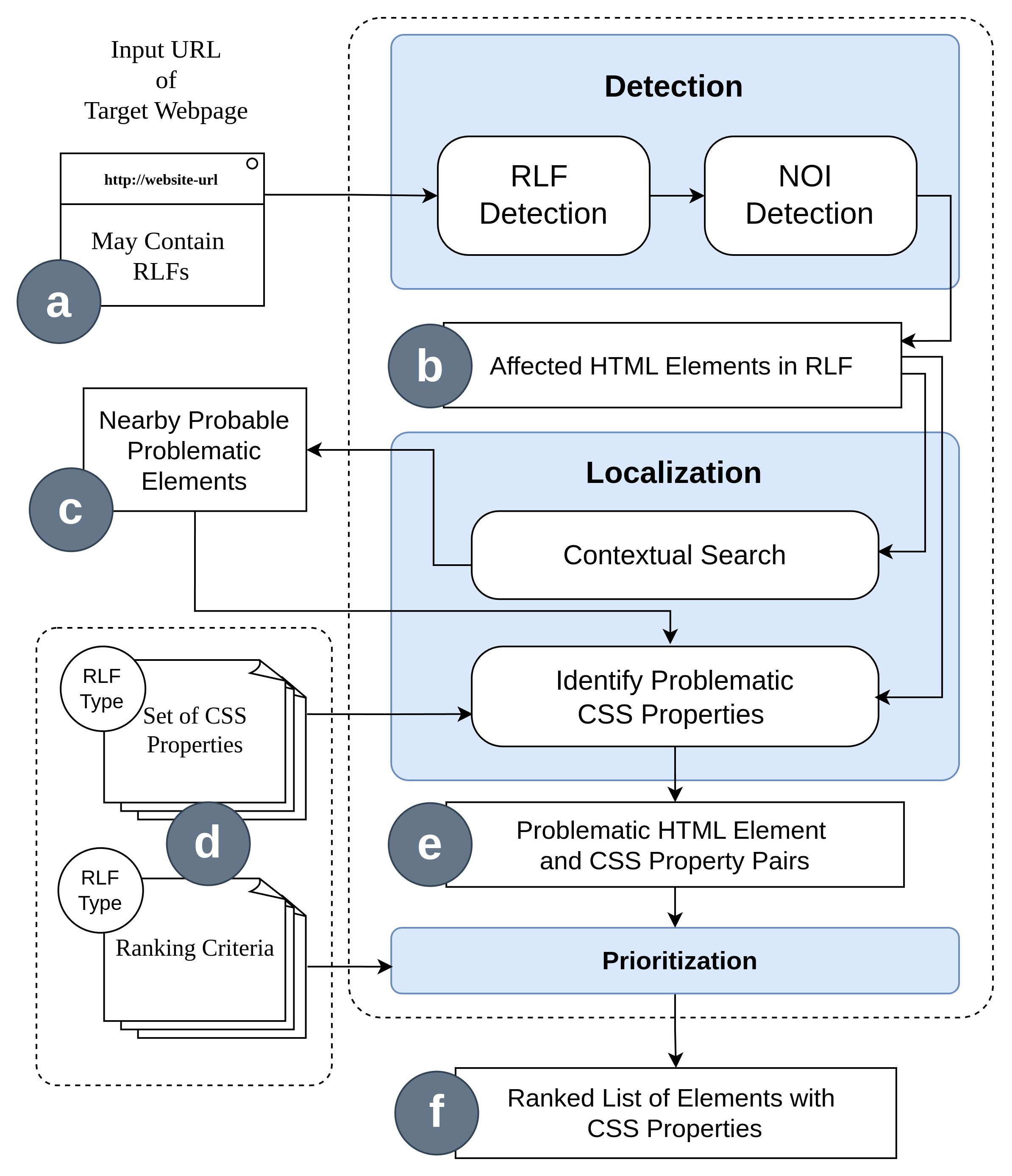}
    \caption{Overview of the Approach}
    \label{fig:High Level Architecture with Example}
\end{figure}

The goal of our approach is to localize HTML elements and corresponding CSS properties responsible for \textit{Responsive Layout Failures} (RLF). The key idea is to develop a heuristic approach that leverages the intuitive reasoning a developer might naturally apply to search for the responsible element properties for a given failure. Identifying the problematic pairs of elements and CSS properties is important for resolving failures both automatically or manually.\\
In a responsively designed webpage, any of the five common types of RLFs \cite{automated_detection} as stated in Section 2 may happen. These RLFs typically occur due to developers adding a wrong value to the properties \cite{automated_detection}. Hence, we solely focus on searching the explicitly defined properties to find the problematic ones.\\
The proposed approach is divided into three distinct phases, \textit{detection}, \textit{localization} and \textit{prioritization}, as shown in Figure \ref{fig:High Level Architecture with Example}. The approach takes a URL (Figure \ref{fig:High Level Architecture with Example}(a)) of a webpage as input. The DOM of this webpage is extracted and sent to Detection phase, where RLFs along with the non-observable issues are detected. This module generates a list of RLFs and affected elements by the failures (Figure \ref{fig:High Level Architecture with Example}(b)). After detection, its output is used in the Localization phase. This module additionally searches for the nearby problematic elements (Figure \ref{fig:High Level Architecture with Example}(c)), and then identifies problematic CSS properties for both - affected elements and nearby elements. To find the properties, a predefined property set for each RLF is used as a reference (Figure \ref{fig:High Level Architecture with Example}(d)). The result is a list for each RLF containing XPath of involved elements with the value of their problematic properties (Figure \ref{fig:High Level Architecture with Example}(e)). Finally, Prioritization phase applies ranking criteria to create a ranked list of these properties (Figure \ref{fig:High Level Architecture with Example}(f)).

\subsection{Phase 1: Detection}
Detection phase is divided into two modules, the first one is to identify the RLFs existing in a webpage. Second one is to detect whether the webpage has any Non-Observable Issues in it. These two modules are explained below:
\subsubsection{RLF Detection}
For this phase, we implemented a detection module same as LAYOUT DR \cite{automated_repair}. This step outputs each of the detected RLFs, comprising its type (one of five RLFs mentioned in Section 2), details about the HTML elements involved in the RLF, and the range of viewports, \texttt{\small $fail_{min}$...$fail_{max}$}, where the failure occurs. Here \texttt{\small $fail_{min}$} is the starting viewport and \texttt{\small $fail_{max}$} is the ending viewport of an RLF.\\
However, elements designed to scroll automatically, such as \texttt{\small carousel, slideshow} change its contents after a specific interval. Elements having such behavior can be detected as a layout change and thus labeled as a failure. To reduce these false positive issues, we additionally identified the presence of CSS \texttt{\small transition} and \texttt{\small transform} properties in each element. Once such an element is found, it is excluded from the search space for failure elements.

\subsubsection{NOI Detection}
RLFs are prone to a special type of false positives known as \textit{Non-Observable Issues (NOIs)} \cite{no_explicit_oracle}, issues that exist in the DOM but have no visible appearance in the page. For instance, one HTML element may overflow the viewable portion of the page, as discovered by checking the element's coordinates in the DOM. However, if it is transparent or contains the same colour as root element (i.e., \texttt{\small <BODY>}), the viewport protrusion will not be visible to a human viewing the page. This type of case is marked as NOI.\\
To determine the NOIs, we implemented VISER, an automated visual verification method \cite{verification}. This method analyzes the failure region of an RLF through image processing by capturing snapshots at various layers. The layers are then compared to find differences in pixel level. If found, the failure is considered observable to a human, otherwise it is non-observable. It allows developers to review NOIs if necessary. However, even VISER is not 100\% accurate, requiring manual checking.\\
The outputs of this phase are the detected RLFs and NOI-marked failures. The affected elements are listed along with each detected RLF.

\subsection{Phase 2: Localization}
Localization phase consists of two modules, the first one searches for nearby problematic elements. Properties of these elements are searched to identify the problematic ones in the second module. These modules are demonstrated below:
\subsubsection{Contextual Search}
A failure region analysis is performed to narrow down the search space of HTML elements and CSS properties. The approach first extracts the coordinates of the failure elements. These coordinates or positions of the elements are used to determine the direction of the RLF. If an RLF is an \textit{element protrusion}, it is essential to determine the direction of the failure—whether it is \textit{horizontal} or \textit{vertical}—and identify the boundary from which the element is protruding (top, bottom, left, or right). Our approach also uses the coordinates of the elements to calculate the relative alignments and identify the relationship between them for a specific range. For instance, Figure \ref{fig:RLFs_comparison}(c) shows a \textit{vertical protrusion}, indicating that any horizontally aligned elements do not have an effect on the RLF. The neighboring elements of a failure are filtered using this alignment information.

\subsubsection{Identify Problematic CSS Properties}
To identify problematic CSS properties of the elements provided by the previous step, a pre-defined set of CSS properties relevant to each RLF type is introduced, denoted as \textit{$P_{type}$}. Properties that control element dimensions (e.g., \texttt{\small height}, \texttt{\small width}) are relevant across all RLF types, whereas properties such as \texttt{\small position} and \texttt{\small display} differ from one type to another. These sets are created by analyzing queries from top two largest and most popular community among developers, \textit{\href{http://www.quora.com}{Quora}} and \textit{\href{http://www.stackoverflow.com}{Stack Overflow}}. RLF-specific terms - \textit{``div overlap"}, \textit{``div overflow"}, \textit{``container overflow"} and \textit{``element wrap"} are used to search for queries related to responsive layout failures. \footnote{\href{https://stackoverflow.com/search?q=div+element+overlap}{StackOverflow - Element Collision}, \href{https://stackoverflow.com/search?q=div+element+overflow}{Element Overflow}, \href{https://stackoverflow.com/search?q=element+going+outside+the+div}{Element Going Outside Div}, \href{https://stackoverflow.com/search?q=element+overflowing+the+page2}{Element Overflowing Page}, \href{https://stackoverflow.com/search?q=elements+wrap+to+a+new+line}{Elements Wrapping}, \href{https://www.quora.com/search?q=div+element+overlap}{Quora - Divs Overlapping}, \href{https://www.quora.com/search?q=div+element+overflow}{Element Overflow}, \href{https://www.quora.com/search?q=element+going+outside+the+div}{Element Going Outside the Div}, \href{https://www.quora.com/search?q=element+overflowing+the+page}{Element Overflowing Page}, \href{https://www.quora.com/search?q=line+up+elements+using+CSS}{Elements Wrapping}}
.\\
Top 20 answers having the highest number of votes are analyzed to identify the involved CSS properties and add them to the sets. Five experienced front-end engineers verified these sets. They manually checked each property of the sets and marked the ones that typically cause an RLF. Properties marked by more than two engineers were taken in the \textit{$P_{type}$} set. However, \textit{small-range} failures do not require any CSS property for localization, as it is dependent on media-query rules. The four other types of RLFs and their corresponding \textit{$P_{type}$} are shown in TABLE \ref{tab:CSS Sets for RLFs}. The \textit{$P_{type}$} is used to search for faulty CSS properties in the elements for a specific RLF. When such a property is found within an element, the approach checks whether it is explicitly defined by the developer. If developer-written, it is added to the output set, \textit{E}, of potentially problematic elements with faulty properties for each RLF. For instance, Figure \ref{fig:RLFs_comparison}(c) shows a \textit{vertical protrusion}, where properties such as \texttt{\small height}, \texttt{\small margin-top}, \texttt{\small margin-bottom}, etc. are potentially involved, while other properties like \texttt{\small margin-left}, \texttt{\small margin-right} are irrelevant.\\

\renewcommand{\arraystretch}{1.1}
\begin{table}[h]
\centering
\caption{Ranked Search Set of CSS Properties for each RLF}
\label{tab:CSS Sets for RLFs}
\resizebox{\columnwidth}{!}{%
\begin{tabular}{l|l}
\hline
Set Type & Ranked CSS Properties \\ \hline
\begin{tabular}[c]{@{}l@{}}Element \\ Protrusion \\ ($P_{EP}$)\end{tabular} & \begin{tabular}[c]{@{}l@{}}1. position: absolute | 2. float | 3. Fixed height, width \\(i.e., px) | 4. display | 5. margin, padding | 6. font-size \\ 7. white-space\end{tabular} \\ \hline
\begin{tabular}[c]{@{}l@{}}Element \\ Collision \\ ($P_{EC}$)\end{tabular} & \begin{tabular}[c]{@{}l@{}}1. position: absolute | 2. float | 3. -ve margin | 4. Fixed \\height, width | 5. margin, padding | 6. If display:flex \\ Missing flex-wrap:wrap | 7. max-height, max-width\end{tabular} \\ \hline
\begin{tabular}[c]{@{}l@{}}Viewport \\ Protrusion \\ ($P_{VP}$)\end{tabular} & \begin{tabular}[c]{@{}l@{}}1. position: absolute | 2. float | 3. Fixed height, width | \\ 4. margin, padding | 5. font-size | 6. white-space\end{tabular} \\ \hline
\begin{tabular}[c]{@{}l@{}}Wrapping \\ Elements \\ ($P_{WE}$)\end{tabular} & \begin{tabular}[c]{@{}l@{}}1. If parent has no display:flex and flex:nowrap | 2. float | \\ 3. parent width | 4. margin, padding | 5. font-size\end{tabular} \\ \hline
\end{tabular}%
}
\vspace{-10pt}
\end{table}

\subsection{Phase 3: Prioritization}
The goal of the last phase is to prioritize set \textit{E}’s elements in order of likelihood to have caused the RLF. It gives a ranked list, $E_{R}$ by sorting each pair (e, c) of set \textit{E}, where e is an HTML element \textit{e $\in$ E} and c is a CSS property \textit{c of e}. Typically, CSS properties assigned with larger numeric values tend to be the reason behind a failure. As a result, sorting by property value is done in descending order.\\ 
For properties having non-numeric value, RLF-specific ranking criteria $RC_{type}$ are used to prioritize the properties. These criteria are based on the properties of set associated with each RLF \textit{$P_{type}$} mentioned in the previous section. Using the obtained answers of queries from Stack Overflow and Quora (mentioned in Sub Section 3.2.2), we took count of the properties mentioned as problematic ones in those answers. TABLE \ref{tab:CSS Sets for RLFs} shows these rankings for each RLF. Specific properties with non-numeric values tend to be the actual root cause of RLFs, e.g., \texttt{\small position}. Hence, they are given the top positions, e.g., in TABLE \ref{tab:CSS Sets for RLFs}, for $P_{EC}$ the $RC_{EC}$ has \texttt{position:absolute} in the 1st rank.\\
If two properties have the same numeric value, $RC_{type}$ will be used to prioritize them. Properties are also prioritized according to the element containing that property. The elements affected by the failure will get the highest priority, followed by their neighboring elements. However, we have not considered the height or width of parent container, as changing its size may distort the layout. The sorted list shows the localized (e, c) pairs in descending order of impact.

\section{\uppercase{Case Study}}
\begin{figure}
    \centering
    \includegraphics[width=1\linewidth]{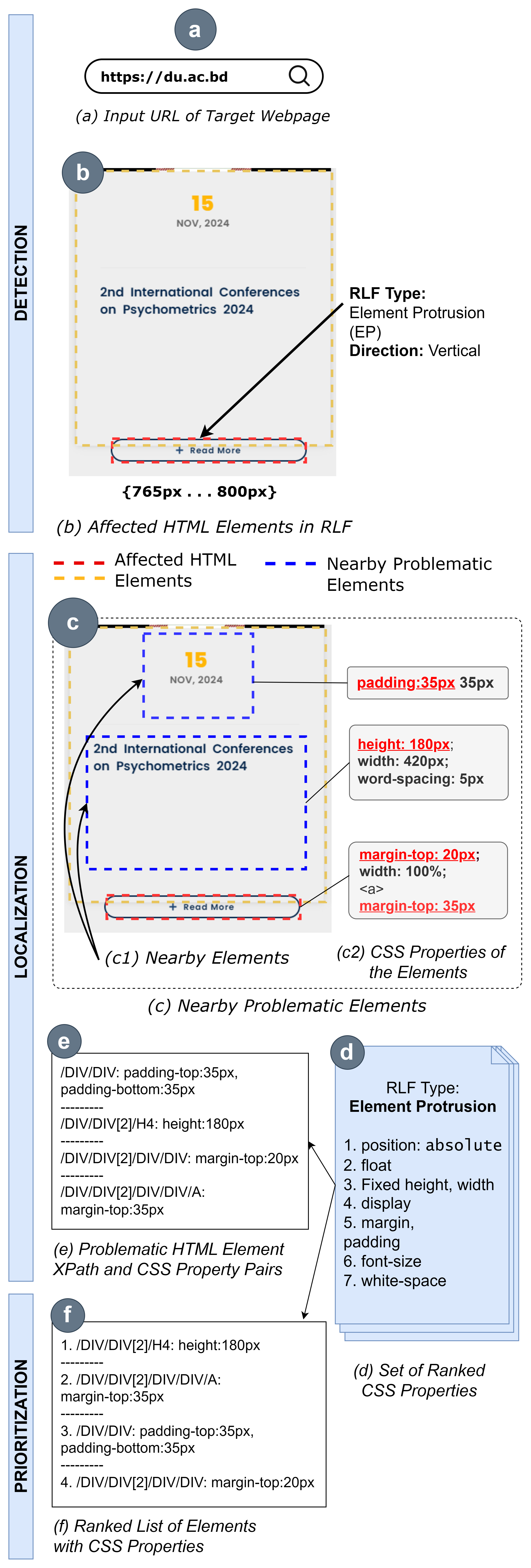}
    \caption{Real-life case study on a webpage demonstrating outputs of different phases by the proposed approach}
    \label{fig:example-showing-outputs}
\end{figure}

Our proposed approach is applied on a sample case using a real-world webpage, and the step by step execution flow is demonstrated here. For this purpose, we have chosen \underline{\href{https:\\du.ac.bd}{du.ac.bd}}, an educational website of a renowned top institution in Bangladesh. We applied our approach to this website and systematically walked through each of the steps. Figure \ref{fig:example-showing-outputs} illustrates the steps and their outputs, using labeled markers to distinguish different outputs.\\ 
\textbf{Detection Phase:} At the beginning, in Figure \ref{fig:example-showing-outputs}(a), the URL of this webpage is given as input. The detection phase takes this webpage, runs through RLF Detection and NOI Detection modules to find any RLFs hidden in the webpage, including NOIs. Here, one RLF is detected, which is an \textit{Element Protrusion}. Its affected elements are pointed in \ref{fig:example-showing-outputs}(b). The dashed button element \textit{``+ Read More"} is overflowing out of the outer-dashed parent container. These elements will act as inputs to the next phase, localization. 
\\
\textbf{Localization Phase:} In this phase, affected elements are used to find their nearby elements which may be problematic. Contextual search is used to identify them by calculating the direction of the RLF. Here, as the button is protruding from the bottom of the container, it can be stated that the RLF direction is vertical. Hence, nearby elements that are vertically aligned can be potentially problematic. In \ref{fig:example-showing-outputs}(c1), the date and title elements are the nearby ones as they are situated on top of the affected button. The CSS properties are also extracted for all these elements in \ref{fig:example-showing-outputs}(c2). To find the problematic properties, only those related to vertical alignment will be considered, such as \texttt{\small margin-top, height, margin-bottom} etc. These properties are underlined in \ref{fig:example-showing-outputs}(c2). Observing the extracted CSS properties will help to understand the process of finding the problematic ones. Here, CSS of the button includes \texttt{\small margin-top, width}, but \texttt{\small width} will not be considered in this case as it does not affect the vertical protrusion.\\
Now localization generates a list consisting of (e, c) pairs, e = element and c = its property, as shown in \ref{fig:example-showing-outputs}(e). The list here contains the underlined properties of \ref{fig:example-showing-outputs}(c2), such as, \texttt{\small margin-top} of the button (Xpaths of the elements are shown), \texttt{\small height} of the element immediate above the button. The pairs are ranked in the next phase.\\
\textbf{Prioritization Phase:} To rank these pairs, the proposed approach utilizes the ranked search set of CSS Properties for each RLF (shown in Table \ref{tab:CSS Sets for RLFs}). Figure \ref{fig:example-showing-outputs}(d) illustrates this list for Element Protrusion. The final output, presented in Figure \ref{fig:example-showing-outputs}(f), shows the $E_{R}$ list, where the properties are sorted in descending order based on their values. In this case, since the \texttt{\small height} property has the highest value within the RLF, it is ranked at the top, followed by \texttt{\small margin-top}, \texttt{\small padding}, and so on. When properties have the same value, the ranked property set from Figure \ref{fig:example-showing-outputs}(d) is used to determine their order. Finally, the list (Figure \ref{fig:example-showing-outputs}(f)) presents the involved elements along with their respective ranks, indicating their level of impact.

\section{\uppercase{Experimental Results}}
We designed an empirical study  to answer the following research questions and evaluate our proposed approach:\\
\textbf{RQ1: How accurately can the proposed approach localize a pair of HTML element and its CSS property for any detected RLF?\\
RQ2: How does the automated localization perform compared to manual localization?}
\\
We carried out the study on a set of 20 real-life webpages containing RLFs. The result is a ranked list of identified (e,c) pairs, where e is an HTML element and c is a CSS property of e. We gave the RLFs to five experienced front-end Engineers for manual localization. Then the results of our approach were compared with the performance of those Engineers.

\subsection{Evaluation Metrics}
The accuracy of the proposed approach is assessed using Top N Rank, Mean Reciprocal Rank (MRR) and Precision at K (P@K), which are widely used for ranking models, such as \cite{bug_localization, bug_localization_noisy}. For all of these, the higher the value is, the better the performance will be. The details of the metrics are given below:\\
\textbf{Top N Rank:} For each RLF, if there is a problematic (e,c) pair within the Top N rank (N = 1, 3, 5, 7 for this system), it is counted as 1. For example, N = 3 means if the (e,c) pair for ${RLF_i}$ is obtained within top 3 suggestions, \textit{isCSSPropertyinTop3(${RLF_i}$)} of Equation \ref{eq:top_k} will be counted as 1. Top N rank is used to evaluate our approach in two ways, (1) if any pair from Top N is the root cause, and (2) if the engineers preferred (e,c) pair exists in Top N of our suggestions.
\begin{equation}
    \small{Top\textrm{-}N = \frac{1}{|RLF_{count}|} \sum_{i}^{|RLF_{count}|} isCSSPropertyinTopN(RLF_i)}
    \label{eq:top_k}
\end{equation}
\\
\textbf{MRR:} A reciprocal rank is the multiplicative inverse of the rank of the first correct result of a query \cite{bug_localization}. For example, if a property is localized in rank position 2, the reciprocal rank is 1/2. MRR is the mean of Reciprocal Ranks across all queries, represented by Equation \ref{eq:MRR}. So, the range of MRR is 0 $\le$ MRR $\le$ 1. Here, $|RLF_{count}|$ denotes the total count of RLFs in a webpage, and $HighestRank_{i}$ refers to the rank of the first correctly identified (e,c) pair for each RLF. MRR measures the effectiveness of identifying the actual root cause of RLFs.

\begin{equation}
    MRR = \frac{1}{|RLF_{count}|} \sum_{i=1}^{|RLF_{count}|} \frac{1}{HighestRank_i}
    \label{eq:MRR}
\end{equation}
\\
$MRR_{total}$ for the whole subject set is calculated using the same Equation (1), where $|RLF_{count}|$ is the total number of detected RLFs by the proposed approach.\\
\textbf{Precision at K (P@K):} It measures the proportion of relevant CSS properties for solving any RLF among the top K retrieved (e,c) pairs. Higher P@K indicates that more relevant results appear within the top K ranks. \(P@K_{i}\) for the \textit{i}th webpage is measured using Equation \ref{eq:precision_at_k}.

\begin{equation}
    \small{P@K_i = \frac{\text{Number of relevant results in the top } K}{K}}
    \label{eq:precision_at_k}
\end{equation}
$P@K_{i}$ is the average of all the individual P@K for each webpages, presented in Equation \ref{eq:precision_at_k}. Here, N is the number of webpages used.
\begin{equation}
P@K = \frac{1}{N} \sum_{i=1}^{N} P@K_i
\label{eq:placeholder}
\end{equation}
\subsection{Experimental Setup and Subjects}
We implemented the proposed approach into a tool called {\normalfont \textsc{LocaliCSS}}\footnote{\href{https://anonymous.4open.science/r/LocaliCSS-4526}{https://anonymous.4open.science/r/LocaliCSS-4526}}. It is implemented with JavaScript using Node v20.11.1. and Puppeteer v20.7.3. In this tool, same as \cite{automated_repair}, we have set the viewport height to 1000 px and width range to 320-1400 px, which varies from small mobile screens to large desktop screen sizes. The tool samples a page at every viewport width within the range, step size = 1 px, similar to \cite{automated_repair}. After setting these configurations, the tool is used to localize the RLFs of subject webpages.\\
\begin{table}[h]
\caption{Subject Webpages}
\label{tab:SUBJECT WEBPAGES USED IN THE EXPERIMENTS}
\resizebox{\linewidth}{!}{%
\begin{tabular}{lllll}
\hline
\textbf{\begin{tabular}[c]{@{}l@{}}Subj \\ ID\end{tabular}} & \textbf{Subject} & \textbf{URL} & \textbf{\begin{tabular}[c]{@{}l@{}}\# Line\\ of HTML\end{tabular}} & \textbf{\begin{tabular}[c]{@{}l@{}}\# Line \\ of CSS\end{tabular}} \\ \hline
1 & 3MinuteJournal & 3minutejournal.com & 79 & 3354 \\
2 & Accountkiller & accountkiller.com/en & 343 & 559 \\
3 & Airbnb & airbnb.com & 1469 & 5638 \\
4 & Ardour & ardour.org & 222 & 3774 \\
5 & Bower & bower.io & 370 & 844 \\
6 & BugMeNot & bugmenot.com & 41 & 237 \\
7 & CoveredCalendar & coveredcalendar.com & 147 & 5131 \\
8 & Cloudconvert & cloudconvert.com & 907 & 2831 \\
9 & Django & djangoproject.com & 242 & 4732 \\
10 & DjangoREST & django-rest-framework.org & 610 & 3787 \\
11 & Duolingo & duolingo.com & 816 & 16929 \\
12 & Honey & joinhoney.com/install & 460 & 3249 \\
13 & HotelWiFiTest & hotelwifitest.com & 358 & 4258 \\
14 & MantisBT & mantisbt.org & 247 & 7731 \\
15 & MidwayMeetup & midwaymeetup.com & 85 & 2942 \\
16 & Ninite & ninite.com & 640 & 2721 \\
17 & PepFeed & pepfeed.com & 342 & 4563 \\
18 & PDFescape & pdfescape.com & 176 & 794 \\
19 & Selenium & selenium.dev & 286 & 4980 \\
20 & WillMyPhoneWork & willmyphonework.net & 781 & 2022 \\ \hline

\end{tabular}%
}
\end{table}
\\
We created our subject set using the webpages studied by \cite{no_explicit_oracle} and \cite{automated_repair}. In total, there were 45 webpages in their online repository. We attempted to run {\normalfont \textsc{LocaliCSS}} on the landing page of those. Among them, pages that did not respond or had no true positive RLFs were discarded. Finally, we found 20 webpages as our subjects, from which 42 distinct failures, including 13 Non-Observable Issues (NOIs) were obtained. TABLE \ref{tab:SUBJECT WEBPAGES USED IN THE EXPERIMENTS} lists these webpages, along with their HTML and CSS line counts. The webpages vary from 41 lines of HTML and 237 lines of CSS to around 1,469 lines of HTML and 16,929 lines of CSS. The variations in their size and complexity helped to form a diverse subject set.

\subsection{Empirical Evaluation}
\textbf{Answer to RQ1.} We ran {\normalfont \textsc{LocaliCSS}} on the subject webpages and found 62 RLFs in total including True Positive (TP) and NOIs. As one webpage can contain the same RLF in different segments, the first author of this paper manually verified the detected RLFs and counted the unique ones. It makes the distinct true positive RLF count 29 and NOI count 13, which are shown in TABLE \ref{Table 1: Top-N Localization results from using the presented approach}(a) including distinct RLF counts for every subject webpage.\\
These RLFs are then localized using the proposed approach, the output contains a ranked list of problematic HTML elements and its CSS property (e,c) pairs. TABLE \ref{Table 1: Top-N Localization results from using the presented approach}(b) shows that {\normalfont \textsc{LocaliCSS}} successfully localized 39 out of 42 RLFs. Other 3 RLFs of Subject ID 9, 14 and 17 respectively were not localized by the tool. Manual investigation revealed that these RLFs have no problematic CSS properties; rather, they are missing a property that is causing RLFs. For instance, in Subject ID 9, an Element Protrusion (EP) is caused by text overflow, which is missing the \texttt{\small line-break} property. The other two RLFs also lack properties, which are out of scope for our tool to detect. Additionally, RLF of Subject ID 8 was not detected by the detection approach we relied on, hence it is missing from 
localized RLFs section.
\renewcommand{\arraystretch}{1.1} % Adjusts row padding
\begin{table*}[t]
\captionsetup{justification=centering}
\caption{Top-N Accuracy Results for LocaliCSS}
\label{Table 1: Top-N Localization results from using the presented approach}
\resizebox{\textwidth}{!}{%
\begin{tabular}{l|cc|cccccc|cccc|c|c|c|c}
\hline
\multicolumn{1}{c|}{\multirow{3}{*}{\textbf{\begin{tabular}[c]{@{}c@{}}Subj\\ ID\end{tabular}}}} & \multicolumn{2}{c|}{\textbf{(a) Detected Failures}} & \multicolumn{6}{c|}{\textbf{(b) Localized Failures}} & \multicolumn{4}{c|}{\textbf{(c) Top-N Accuracy}} & \multirow{4}{*}{\textbf{\begin{tabular}[c]{@{}c@{}}(d) \\ MRR (\%)\end{tabular}}} & \multirow{4}{*}{\textbf{\begin{tabular}[c]{@{}c@{}}(e) \\ MRR \\ w/o WE\\ (\%)\end{tabular}}} & \multirow{4}{*}{\textbf{\begin{tabular}[c]{@{}c@{}}(f) \\ P@3\end{tabular}}} & \multirow{4}{*}{\textbf{\begin{tabular}[c]{@{}c@{}}(g) \\ P@3\\ w/o WE\end{tabular}}} \\ \cline{2-13}
\multicolumn{1}{c|}{} & \multicolumn{2}{c|}{\textbf{Total Distinct RLF(s)}} & \multicolumn{1}{l|}{\multirow{2}{*}{\textbf{EP}}} & \multicolumn{1}{l|}{\multirow{2}{*}{\textbf{EC}}} & \multicolumn{1}{l|}{\multirow{2}{*}{\textbf{VP}}} & \multicolumn{1}{l|}{\multirow{2}{*}{\textbf{WE}}} & \multicolumn{1}{l|}{\multirow{2}{*}{\textbf{SR}}} & \multirow{2}{*}{\textbf{\begin{tabular}[c]{@{}c@{}}Total \\ Distinct \\ RLF(s)\end{tabular}}} & \multicolumn{1}{c|}{\multirow{3}{*}{\textbf{Top-1}}} & \multicolumn{1}{c|}{\multirow{3}{*}{\textbf{Top-3}}} & \multicolumn{1}{c|}{\multirow{3}{*}{\textbf{Top-5}}} & \multirow{3}{*}{\textbf{Top-7}} &  &  &  &  \\ \cline{2-3}
\multicolumn{1}{c|}{} & \multirow{2}{*}{TP} & \multirow{2}{*}{NOI} & \multicolumn{1}{l|}{} & \multicolumn{1}{l|}{} & \multicolumn{1}{l|}{} & \multicolumn{1}{l|}{} & \multicolumn{1}{l|}{} &  & \multicolumn{1}{c|}{} & \multicolumn{1}{c|}{} & \multicolumn{1}{c|}{} &  &  &  &  &  \\
 &  &  & \multicolumn{1}{l|}{} & \multicolumn{1}{l|}{} & \multicolumn{1}{l|}{} & \multicolumn{1}{l|}{} & \multicolumn{1}{l|}{} & \multicolumn{1}{l|}{} & \multicolumn{1}{c|}{} & \multicolumn{1}{c|}{} & \multicolumn{1}{c|}{} &  &  &  &  &  \\ \hline
1 & \textbf{1} & \textbf{3} & 3 & - & 1 & - & - & \textbf{4} & 3 & 4 & 4 & 4 & 87.50 & 87.50 & 0.92 & 0.92 \\
2 & \textbf{1} & \textbf{-} & - & - & - & - & 1 & \textbf{1} & 1 & 1 & 1 & 1 & 100 & 100 & 1 & 1 \\
3 & \textbf{2} & \textbf{-} & - & - & - & 2 & - & \textbf{2} & - & 2 & 2 & 2 & 50 & - & 0.67 & - \\
4 & \textbf{2} & \textbf{-} & 1 & - & 1 & - & - & \textbf{2} & - & 2 & 2 & 2 & 50 & 50 & 0.5 & 0.5 \\
5 & \textbf{3} & \textbf{-} & 1 & 1 & 1 & - & - & \textbf{3} & 2 & 3 & 3 & 3 & 77.78 & 77.78 & 0.89 & 0.89 \\
6 & \textbf{2} & \textbf{-} & 1 & - & - & 1 & - & \textbf{2} & 1 & 1 & 2 & 2 & 62.50 & 62.50 & 0.5 & 0.5 \\
7 & \textbf{1} & \textbf{-} & - & - & - & 1 & - & \textbf{1} & - & - & 1 & 1 & 20 & - & 0 & - \\
8 & \textbf{0} & \textbf{-} & - & - & - & - & - & \textbf{0} & - & - & - & - & - & - & - & - \\
9 & \textbf{2} & \textbf{2} & 2 & - & - & 1 & - & \textbf{3} & 2 & 3 & 3 & 3 & 62.50 & 66.67 & 1 & 1 \\
10 & \textbf{1} & \textbf{-} & - & - & 1 & - & - & \textbf{1} & - & 1 & 1 & 1 & 50 & 50 & 0.67 & 0.67 \\
11 & \textbf{2} & \textbf{-} & - & - & - & 1 & 1 & \textbf{2} & 1 & 2 & 2 & 2 & 75 & 100 & 0.67 & 0.67 \\
12 & \textbf{1} & \textbf{-} & - & - & - & 1 & - & \textbf{1} & - & - & 1 & 1 & 20 & - & 0 & - \\
13 & \textbf{1} & \textbf{-} & - & - & 1 & - & - & \textbf{1} & 1 & 1 & 1 & 1 & 100 & 100 & 1 & 1 \\
14 & \textbf{1} & \textbf{3} & 2 & - & - & 1 & - & \textbf{3} & 1 & 2 & 3 & 3 & 43.75 & 50 & 0.44 & 0.41 \\
15 & \textbf{1} & \textbf{-} & 1 & - & - & - & - & \textbf{1} & 1 & 1 & 1 & 1 & 100 & 100 & 1 & 1 \\
16 & \textbf{1} & \textbf{-} & - & - & - & 1 & - & \textbf{1} & - & 1 & 1 & 1 & 50 & - & 0.67 & - \\
17 & \textbf{2} & \textbf{2} & 1 & 1 & 1 & - & - & \textbf{3} & - & 2 & 3 & 3 & 31.25 & 31.25 & 0.56 & 0.56 \\
18 & \textbf{3} & \textbf{-} & 1 & - & 2 & - & - & \textbf{3} & 2 & 2 & 3 & 3 & 75 & 75 & 0.67 & 0.67 \\
19 & \textbf{1} & \textbf{3} & 2 & 1 & - & 1 & - & \textbf{4} & 3 & 3 & 3 & 4 & 79.17 & 79.17 & 0.75 & 0.75 \\
20 & \textbf{1} & \textbf{-} & 1 & - & - & - & - & \textbf{1} & 1 & 1 & 1 & 1 & 100 & 100 & 1 & 1 \\ \hline
\textbf{Total} & \textbf{29} & \textbf{13} & \textbf{16} & \textbf{3} & \textbf{8} & \textbf{10} & \textbf{2} & \textbf{39} & \textbf{\begin{tabular}[c]{@{}c@{}}19\\ (45.2\%)\end{tabular}} & \textbf{\begin{tabular}[c]{@{}c@{}}32\\ (76.2\%)\end{tabular}} & \textbf{\begin{tabular}[c]{@{}c@{}}38\\ (90.5\%)\end{tabular}} & \textbf{\begin{tabular}[c]{@{}c@{}}39\\ (92.86\%)\end{tabular}} & \textbf{68\%} & \textbf{76\%} & \textbf{\begin{tabular}[c]{@{}c@{}}0.66\\ (66.67\%)\end{tabular}} & \textbf{\begin{tabular}[c]{@{}c@{}}0.77\\ (77.13\%)\end{tabular}} \\ \hline
\end{tabular}%
}
\begin{minipage}{\textwidth}
\vspace{0.1cm}
\parbox{\columnwidth}{\footnotesize \textit{Note:} Complete detection result can be viewed in LocaliCSS\footnotemark[1] repository}
\end{minipage}
\end{table*}
\\
We manually inspected each generated (e, c) pair to verify that (1) it can remove the original RLF if we modify that CSS property (c), (2) it does not introduce more RLFs when the detection module is rerun. These two rules are considered as ground truth for our first evaluation. If a pair meets both criteria, it is categorized as Top-1, Top-3, Top-5 or Top-7 based on its position in the generated ranked list. In TABLE \ref{Table 1: Top-N Localization results from using the presented approach}(c), Top-1 indicates the first pair in the generated list is the problematic pair for 45.2\% of 42 RLFs. In 76.19\% of the cases, one or all of the first three pairs are likely the root cause of the RLF. Similarly, Top-5 and Top-7 both achieve accuracies of 90.5\% and 92.86\%, respectively.\\
MRR for each of the Subject webpages is also calculated and shown in TABLE \ref{Table 1: Top-N Localization results from using the presented approach}(d). The results show that $MRR_{total}$ of {\normalfont \textsc{LocaliCSS}} is 68\%, indicating that the problematic pair is, on average, ranked among top 2 in the generated list. Among the 20 subject webpages, 14 webpages achieved an MRR of at least 50\%, indicating significant localization performance. Remaining webpages had an MRR between 20\% and 50\%, indicating the correct localization appeared lower in the ranked list. Through manual investigation of those webpages, it was found that webpages having wrapping failure were not localized properly. Some localized properties were distorting the layout of the wrapped elements, resulting in no proper solution and thus not localizable. Manual localization by our engineers also confirmed these cases as non RLFs. After excluding such cases, we achieved an MRR of a minimum 50\% for all webpages except Subject ID 17, as shown in TABLE \ref{Table 1: Top-N Localization results from using the presented approach}(e). On the contrary, for Subject ID 17, MRR is 31\% due to the developer's use of unnecessary \texttt{float} and \texttt{position} properties, making layout adjustments more challenging.\\
P@K is measured in the similar way as MRR, shown in TABLE \ref{Table 1: Top-N Localization results from using the presented approach}(f) and (g). K=3 is optimal for this measurement, as  K=1 only considers the top result like Top-1 and K=5, K=7 includes too many results, making it less precise. Here P@3 for {\normalfont \textsc{LocaliCSS}} is 66.67\%, showing that for any detected RLF, our approach provides relevant properties in the first 3 positions 66.67\% of the time. As some Wrapping Element failures are not considered RLF by the engineers, excluding those failures increases the P@3 to 77.13\%. P@3 measures how many of the first 3 properties from the output can fix an RLF. Suppose for an RLF, P@3 = 1.0 (100\%), it means the RLF can be repaired by using any of those three properties. Thus P@3 shows the proportion of relevant (e,c) pairs generated by our approach for any given RLF.

\renewcommand{\arraystretch}{1.2} % Adjusts row padding
\begin{table*}[t]
\captionsetup{justification=centering}
\caption{Top-N Accuracy Compared to Manual Localization}
\label{Tab: Localization Results for Developers Vs Tool new}
\centering
\resizebox{0.85\textwidth}{!}{%
\begin{tabular}{l|cccccc|cccc}
\hline
\multirow{2}{*}{\textbf{\begin{tabular}[c]{@{}l@{}}Subject\\ ID\end{tabular}}} & \multicolumn{6}{c|}{\textbf{(a) Manual Localization}} & \multicolumn{4}{c}{\textbf{(b) Top-N Accuracy}} \\ \cline{2-11} 
 & \multicolumn{1}{l}{\textbf{EP}} & \multicolumn{1}{l}{\textbf{EC}} & \textbf{VP} & \multicolumn{1}{l}{\textbf{WE}} & \multicolumn{1}{l}{\textbf{SR}} & \textbf{\begin{tabular}[c]{@{}c@{}}Total Distinct RLF(s)\end{tabular}} & \textbf{Top-1} & \textbf{Top-3} & \textbf{Top-5} & \textbf{Top-7} \\ \hline
1 & 3 & - & 1 & - & - & \textbf{4} & \textbf{3} & \textbf{4} & \textbf{4} & \textbf{4} \\
2 & - & - & - & - & 1 & \textbf{1} & \textbf{1} & \textbf{1} & \textbf{1} & \textbf{1} \\
3 & - & - & - & 2 NP & - & \textbf{2} & \textbf{-} & \textbf{2} & \textbf{2} & \textbf{2} \\
4 & 1 & - & 1 & - & - & \textbf{2} & \textbf{-} & \textbf{2} & \textbf{2} & \textbf{2} \\
5 & 1 & 1 & 1 & - & - & \textbf{3} & \textbf{2} & \textbf{3} & \textbf{3} & \textbf{3} \\
6 & 1 & - & - & 1 & - & \textbf{2} & \textbf{1} & \textbf{1} & \textbf{2} & \textbf{2} \\
7 & - & - & - & 1 NP & - & \textbf{1} & \textbf{-} & \textbf{-} & \textbf{1} & \textbf{1} \\
8 & - & 1 & - & - & - & \textbf{1} & \textbf{-} & \textbf{-} & \textbf{-} & \textbf{-} \\
9 & 3 & - & - & 1 NP & - & \textbf{4} & \textbf{2} & \textbf{3} & \textbf{3} & \textbf{3} \\
10 & - & - & 1 & - & - & \textbf{1} & \textbf{-} & \textbf{1} & \textbf{1} & \textbf{1} \\
11 & - & - & - & 1 NP & 1 & \textbf{2} & \textbf{1} & \textbf{2} & \textbf{2} & \textbf{2} \\
12 & - & - & - & 1 NP & - & \textbf{1} & \textbf{-} & \textbf{-} & \textbf{1} & \textbf{1} \\
13 & - & - & 1 & - & - & \textbf{1} & \textbf{1} & \textbf{1} & \textbf{1} & \textbf{1} \\
14 & 3 & - & - & 1 NP & - & \textbf{4} & \textbf{1} & \textbf{2} & \textbf{3} & \textbf{3} \\
15 & 1 & - & - & - & - & \textbf{1} & \textbf{1} & \textbf{1} & \textbf{1} & \textbf{1} \\
16 & - & - & - & 1 NP & - & \textbf{1} & \textbf{-} & \textbf{1} & \textbf{1} & \textbf{1} \\
17 & 1 & 2 & 1 & - & - & \textbf{4} & \textbf{-} & \textbf{2} & \textbf{3} & \textbf{3} \\
18 & 1 & - & 1 (1 NP) & - & - & \textbf{3} & \textbf{2} & \textbf{2} & \textbf{3} & \textbf{3} \\
19 & 2 & 1 & - & 1 & - & \textbf{4} & \textbf{2} & \textbf{2} & \textbf{3} & \textbf{3} \\
20 & 1 & - & - & - & - & \textbf{1} & \textbf{1} & \textbf{1} & \textbf{1} & \textbf{1} \\ \hline
\textbf{Total} & \textbf{18} & \textbf{5} & \textbf{8} & \textbf{10} & \textbf{2} & \textbf{43} & \textbf{\begin{tabular}[c]{@{}c@{}}18\\ (42.86\%)\end{tabular}} & \textbf{\begin{tabular}[c]{@{}c@{}}31\\ (73.81\%)\end{tabular}} & \textbf{\begin{tabular}[c]{@{}c@{}}38 \\ (90.48\%)\end{tabular}} & \textbf{\begin{tabular}[c]{@{}c@{}}38 \\ (90.48\%)\end{tabular}} \\ \hline
\end{tabular}%
}

\end{table*}
\noindent

\textbf{Answer to RQ2.} We asked five front-end engineers, each with two years of experience, to manually verify the detected RLFs and localize their problematic properties. They verified all 42 RLFs as TPs and performed localization for them. These localized properties for each RLF are considered as ground truth for this evaluation. \\
TABLE \ref{Tab: Localization Results for Developers Vs Tool new}(a) shows these results for all RLFs. Since engineers may suggest different (e,c) pairs, we took opinions from two engineers for each RLF. We used that if both suggested (e, c)s were the same. Otherwise, took an opinion from a third engineer. If that differed too, we chose one from three of them. Five cases required a third opinion, and only one was resolved based on our decision. The engineers marked 9 failures as NP (No Problem). These detected issues were not considered failures by them. For instance, when there is not enough horizontal space to fit every icon side-by-side in the footer, they wrap to a new line, triggering \textit{wrapping RLF}. However, this behavior is typically intentional by developers to fit icons on small screens. One RLF from Subject ID 8 was manually detected but not identified by the detection approach we relied on. This makes one additional RLF manually localized.\\
TABLE \ref{Tab: Localization Results for Developers Vs Tool new}(b) shows Top N accuracy results for the effectiveness of our approach. We searched manually localized $(e,c)_{manual}$ pairs in our generated ranked list. If $(e,c)_{manual}$ is same as the first one $(e,c)_{generated}$ in the list, it is categorized as Top-1, similarly Top-3, Top-5 and Top-7 are also calculated. However, for 4 RLFs of Subject ID 9, 14, 17 and 19 respectively, properties selected by engineers did not appear in the generated list. The reasons for cases of Subject ID 9, 14 and 17 as explained earlier. They lack an additional property which the engineers identified, but localization of these properties is beyond the tool’s scope. For subject ID 19, our tool correctly localized 4 RLFs, of which two are categorized as Top-1 and Top-3. The 3rd RLF is caused by \texttt{\small padding} of a parent and its child. Engineers, with more contextual knowledge, may know which one to give preference. In this, {\normalfont \textsc{LocaliCSS}} suggested modifying the parent’s \texttt{\small padding} for its larger value, while engineers preferred adjusting the child’s \texttt{\small padding} — both are valid choices. This $(e,c)_{generated}$ was ranked 5th in our generated list, causing this pair to be in Top-5. For the 4th RLF, \textit{wrapping element}, the tool gave priority on changing the large \texttt{\small width} value of the wrapped item. On the other hand, engineers suggested changing the small \texttt{\small margin} value for each item of the row, which is a more effective solution than the automated one. However, properties from the generated list can also give a fix.\\
In case of the No Problem (NP) marked failures, we took Top N results of those from TABLE \ref{Table 1: Top-N Localization results from using the presented approach}(c) since the tool was able to localize them. We calculated our result by excluding the RLF of Subject ID 8, as it was not detected by {\normalfont \textsc{LocaliCSS}}. This makes a total of 42 RLFs. Among these, the tool successfully found the problematic (e, c) pairs of 18 cases at the first one, making Top-1 accuracy 42.86\%. Top-3 accuracy is 73.81\%, which is also satisfactory. Finally, Top-5 and Top-7 show that {\normalfont \textsc{LocaliCSS}} can localize 90.48\% accurately as developers do.\\

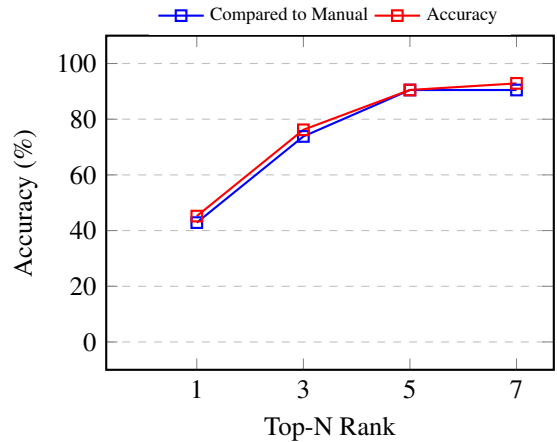
\begin{figure}[h]
    \centering
    \begin{tikzpicture}
        \begin{axis}[
            xlabel={Top-N Rank},
            ylabel={Accuracy (\%)},
            width=1\linewidth,  % Adjusted width for two-column papers
            height=6cm,
            xmin=0, xmax=7,
            ymin=0, ymax=100,
            xtick={1,3,5,7},
            ytick={0,20,40,60,80,100},
            ymajorgrids=true,
            grid style=dashed,
            draw=black,  % Adds an outer border
            thick,
            enlargelimits=true,
            % legend pos=outer north,
            legend style={at={(0.5,1)}, draw=none, anchor=south, scale=0.8, legend columns=-1, align=left}
        ]

        % Plot: Compared to Manual Localization (Blue)
        \addplot[
            color=blue,
            mark=square,
            thick
        ]
        coordinates {
            (1,42.86) (3,73.81) (5,90.48) (7,90.48)
        };
        \addlegendentry{Compared to Manual}

        % Plot: Accuracy (Red)
        \addplot[
            color=red,
            mark=square,
            thick
        ]
        coordinates {
            (1,45.2) (3,76.2) (5,90.5) (7,92.86)
        };
        \addlegendentry{Accuracy}

        \end{axis}
    \end{tikzpicture}
    \caption{Top-N Accuracy Comparison for Automated and Manual Localization}
    \label{fig:topN_accuracy}
\end{figure}
Figure 4 holds a comparative study between the results of automated and manual localization using two different metrics. The Top-N ranking for the accuracy of our approach is similar to that compared to manual localization. It indicates that the first correct CSS property identified by our approach is also the one proposed by the engineers. As N increases, the accuracy of our approach continues to improve, closely following the manual localization. By Top-5 and Top-7, the results converge, reaching around 90-95\% accuracy. This similarity proves that our approach successfully identifies problematic CSS properties in a way that matches how developers manually find them.

\subsection{Threats to Validity}
One threat to the validity of the results is the generalizability of the webpages used in our study. To address this concern, we selected webpages of varying sizes and types, as shown in Table \ref{tab:SUBJECT WEBPAGES USED IN THE EXPERIMENTS}. This approach is consistent with prior studies \cite{automated_repair, no_explicit_oracle}, which also focused on diverse webpage samples.\\ Another possible threat comes from the detection module, developed as Layout DR \cite{automated_repair}, which identifies RLFs. Since we rely on this module for detection, its accuracy directly influences the results. The ranked search set of CSS properties (Table \ref{tab:CSS Sets for RLFs}) was derived from the top 20 queries collected from two prominent platforms, verified by experts. In the future, this set can be expanded by incorporating a broader range of queries and additional platforms. While multiple keywords were used to retrieve the relevant queries, increasing the variety of keywords could strengthen this derivation process.\\
Finally, subjectivity remains a potential threat when manually localizing RLFs. To reduce this, we consulted two engineers for each RLF localization, and in cases of disagreement, we took a third opinion to ensure accuracy.

\section{\uppercase{Related Work}}
Research on detecting and repairing \textit{Responsive Layout Failures} (RLFs) has not yet been extensively explored. Most of the work focuses on detecting RLFs, while comparatively less focus on automated approaches to repair them.\\
For webpages, Walsh et al. \cite{automated_detection} first introduced a method to automatically detect RLFs. They implemented a graph called \textit{Responsive Layout Graph} (RLG), which has HTML elements as nodes and their relationships as edges. To identify changes, their method compares between RLGs of previous and current version of a webpage. They also developed a tool called {\normalfont \textsc{ReDeCheck}} \cite{redecheck} for this purpose. However, the limitation of this approach is requiring a previous version available for a webpage to detect regressions.\\
Addressing this issue, Walsh et al. \cite{no_explicit_oracle} later presented another version of {\normalfont \textsc{ReDeCheck}} to check the layout of a responsive web page against itself. It uses the RLG to search for changing alignments and relationships among the elements. It compares the relative positioning of elements at different viewport widths. Thus, it does not require any explicit oracle, e.g., mockup images or a graph model of the page to compare against. The study introduced five types of RLFs prevalent in real-world web pages. These RLF types have been extensively used in the following works, including this paper. They are, \textit{Element Collision, Element Protrusion, Viewport Protrusion, Small-Range} and \textit{Wrapping Elements}. Section 2 shows these types with their definitions. Although {\normalfont \textsc{ReDeCheck}} reliably detects these RLFs in a responsive page, it does not provide an automatic repair of a reported RLF.\\
To repair RLFs, Jacquet et al. \cite{linear_programming} presented an approach using linear programming to repair such layout failures. Their technique requires a developer to set constraint specifications of the desired layout of a webpage. Moreover, it only works with a fixed viewport of any webpage, hence not suitable for responsive webpages. Later on, Althomali et al. \cite{automated_repair} proposed a tool called {\normalfont \textsc{Layout DR}}, which sources layouts from either side of the affected viewport range, free from RLFs. The layouts are scaled and transformed within the failure range to create two potential \textit{“hotfixes”}. The entire layout is modified in this process, which fixes the affected element but causes a change to almost every other property value explicitly defined by developers. However, localizing and addressing only the CSS properties can be a more suitable option.\\
Apart from RLFs, detection and localization approaches have been used for other web application failures. For instance, ``Alignment Graph” (AG) \cite{XPERT} models layout for detecting cross-browser issues (XBIs). Mahajan et al. used search-based \cite{IFIX, repair_internationalization_search_based_techniques} techniques to detect presentational failures and localize HTML elements responsible for the failure. Later on, Mahajan et al \cite{mob-friendly-problems} worked on mobile friendly problems, such as \textit{font-size, tap-target} etc. They introduced an automated method to localize CSS and generate patches to solve those issues of a web page. Later on, a better approach was proposed by \cite{aesthetics}, which was an automated repair approach for mobile friendly problems based on meta-heuristic algorithms. This approach assured both usability and aesthetics. Zhe Liu et al. presented ``Nighthawk" \cite{nighthawk_localize_visual_understanding} to detect GUIs with display issues and locate region of the issue for guiding developers to fix the bug. Research on web accessibility also uses detection and localization approaches, such as \cite{reflow_accessibility}, a novel technique to automatically detect reflow accessibility issues and localize the failure elements in web pages for keyboard users. Apart from responsiveness, numerous techniques have been developed to evaluate webpages from various perspectives, including the automation of web testing through different approaches \cite{compare_automation, web_mutation_testing, integrate_ai}.\\
All of the techniques discussed in this section tackle specific types of webpage failures. None of them, however, worked on localizing RLFs as does this paper. Another difference is, existing repair approaches still lack complete usability as they are a hotfix, not a permanent solution. Generating a repair to be used as a patch and sent to production is yet to be done. Thus, {\normalfont \textsc{LocaliCSS}}, presented in this paper, attempts to enhance the existing state-of-the-art work on web page repair by doing its prior step, localization. 

\section{\uppercase{Conclusion and Future Work}}
The paper introduces an approach to heuristically search for HTML elements that are potentially responsible to cause an RLF. The involved CSS properties are then localized. By using the ranked properties set for each RLF, it prioritizes the potential properties based on their value and rank. The approach outputs a set of elements and their properties that have the most impact on an RLF.\\
Our approach localized the correct properties with an accuracy of 45.2\% to 92.86\% for Top-1 to Top-7 respectively. The approach also achieves an MRR of 76\%, indicating that the correct result is achieved among the top 2 properties of the generated list. Additionally, P@3 is 77.13\%, meaning that the top 3 suggested properties are relevant to the developers 77.13\% of the time. The comparison with manual localization revealed that our suggested list contains their preferred choices in Top-1 to Top-7 by  42.86\% to 90.48\% respectively. Overall, these results indicate that our approach can assist developers in localizing any RLF, reducing the time and manual effort required, especially for complex webpage designs. By identifying the relevant CSS properties, developers can quickly verify and address the localized issues, making the RLF repair process easier. \\
In future, the research can be directed to decrease differences between manual and automated localization. Increasing the Top-1 accuracy of our approach will match the suggested properties with the developers preferences more accurately. Better localization will play a significant role in generating a more suitable repair solution for the RLFs.

\section*{\uppercase{Acknowledgements}}
This research is supported by the Fellowship from ICT Division, Government of Bangladesh; No-56.00.0000.052.33.002.24-61, dated 06.06.2024.

% \clearpage

\bibliographystyle{apalike}
{\small
\bibliography{references}}

\end{document}